\documentclass[letterpaper]{jpconf}
\usepackage{graphicx}
\usepackage{amssymb}
\usepackage{amsmath}

\begin{document}
\title{Blind analysis results of the TWIST experiment}

\author{Anthony Hillairet, for the TWIST Collaboration%
\protect\footnote{~~\texttt{http://twist.triumf.ca/}: 
R. Bayes, Yu.I. Davydov, W. Faszer, M.C. Fujiwara, A. Grossheim,
D.R. Gill, P. Gumplinger, A. Hillairet, R.S. Henderson, J. Hu,
G.M. Marshall, R.E. Mischke, K. Olchanski, A. Olin,
R. Openshaw, J.-M. Poutissou, R. Poutissou,
G. Sheffer, B. Shin (TRIUMF), 
A. Gaponenko, R.P. MacDonald (University of Alberta), 
J.F. Bueno, M.D. Hasinoff (University of British Columbia), 
P. Depommier (Universit\'e de Montr\'eal), 
E.W. Mathie, R. Tacik (University of Regina), 
V. Selivanov (Russian Research Center, Kurchatov Institute), 
C.A. Gagliardi, R.E. Tribble, (Texas A\&M),
D.D. Koetke, T.D.S. Stanislaus (Valparaiso University).
 }}

\address{University of Victoria, Victoria, Canada}

\begin{abstract}
The TRIUMF Weak Interaction Symmetry Test (TWIST) experiment was designed
to test the standard model at high precision in the purely leptonic decay 
of polarized muons. A general four-fermion interaction model is used to describe
the muon decay. TWIST measures three of the four
muon decay parameters of this model, $\rho$, $\delta$ and $P_{\mu}^{\pi} \xi$,
from the shape of the momentum-angle spectrum. The results of this model independent approach
are compared to the standard model predictions and used to constrain new physics.

Our collaboration has finalized the blind analysis of the final experimental data
taken in 2006 and 2007. This analysis mostly reached our goal of a precision of
an order of magnitude improvement over the pre-TWIST measurements.

\end{abstract}

\section{Introduction}
The discovery of new physics is expected at the high energies reached
by the LHC. However low energy physics such as muon decay
can also play a complementary role in probing
physics beyond the standard model (SM).

Muon decay is a purely leptonic process that is ideal for testing the weak interaction
at high precision. A model independent approach is possible due to the large
mass of the W boson compared to the muon mass.
The most general
Lorentz-invariant, derivative-free, lepton-number-conserving
matrix element $M$ describing muon decay can be written
in terms of helicity-preserving amplitudes as  \cite{Fetscher:1986}
\begin{equation}
  M = \frac{4 G_F}{\sqrt{2}}
   \sum_{\substack{i=L,R\\j=L,R\\ \kappa=S,V,T}} g_{ij}^\kappa 
   \bigl\langle\bar\psi_{e_i} \bigl\vert\Gamma^\kappa\bigr\vert
   \psi_{\nu_e}\bigr\rangle
   \bigl\langle\bar\psi_{\nu_\mu}
   \bigl\vert\Gamma_\kappa\bigr\vert \psi_{\mu_j}\bigr\rangle,
  \label{eq:mudecay_matrixelem}
\end{equation}
where $g_{ij}^{\kappa}$ are the complex weak coupling constants
and $\Gamma^\kappa$ are the possible interactions
(scalar, vector, tensor). In this notation, the SM
postulates that $g_{LL}^{V} = 1$, and $g_{ij}^{\kappa}=0$ otherwise.
If the polarization of the decay positron is undetected,
then the differential decay rate can be expressed as
\begin{equation}
  \frac{d^2\Gamma}{dx \, d\cos\theta}
   \propto
   F_{IS} (x) + P_\mu \xi\cos\theta \, F_{AS} (x),
\label{e:differential_decay_rate}
\end{equation}
where $x=E_e/E_{\textrm{max.}}$, $E_{\textrm{max.}}$ is the maximum energy of the positron,
$\theta$ is the angle between the muon polarization and the positron
momentum,
$P_\mu = |\vec{P}_\mu|$ (the degree of muon polarization), and
\begin{eqnarray}
  &F_{IS}(x)  =  x(1-x) + \frac{2}{9} \rho \left( 4x^2 - 3x - x_0^2 \right) + \eta x_0 (1-x) + \text{R.C.,} \label{e:Fisotropic} \\
  &F_{AS}(x)  =  \frac{1}{3} \sqrt{x^2-x_0^2} \left[ 1 - x + \frac{2}{3} \delta \left( 4x - 3 + \left( \sqrt{1-x_0^2} - 1 \right) \right) \right] + \text{R.C.}\label{e:Fanisotropic}
\end{eqnarray}
The R.C. terms are radiative corrections, which become
more significant as $x$ approaches one. The dimensionless electron mass is
defined by $x_0 = {m_e}/E_{\textrm{max.}}$.
The muon decay parameters $\rho$, $\delta$, $\xi$ and $\eta$ are bilinear
combinations of the weak coupling constants $g_{ij}$.
The TWIST experiment measures $\rho$, $\delta$ and $P_\mu^\pi \xi$ to
parts in $10^4$ from the momentum-angle of the decay positron. The polarization of the muon
from pion decay is $P_{\mu}^{\pi}$. The SM predicts that $\rho=\delta=3/4$, $P_{\mu}^{\pi}=\xi=1$,
and $\eta = 0$; deviations from these predictions would indicate
new physics.

\section{Experiment}

\begin{figure}
\centering
  \includegraphics[height=.3\textheight]{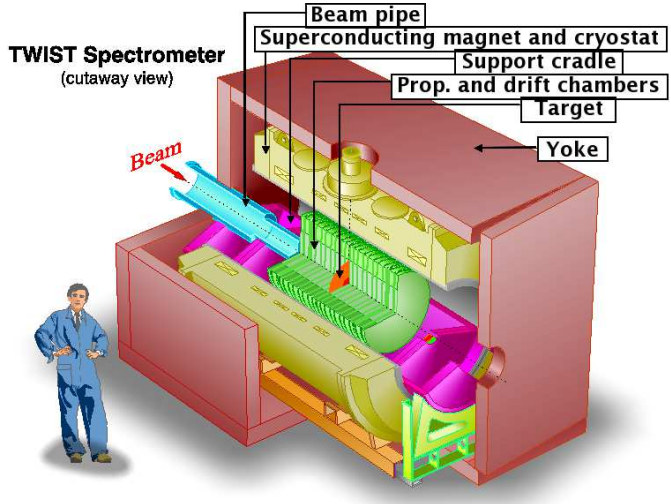}
  \begin{minipage}[b]{14pc}
  \caption{\label{f:twist_spectrometer}The TWIST spectrometer. A highly polarized muon beam is stopped in a thin metal target.
The muons decay to positrons that are tracked in a 2 T magnetic field by an array of planar drift chambers.}
  \end{minipage}
\end{figure}
The experiment used the M13 beam line at TRIUMF in Vancouver, Canada.
Positive pions decaying at rest at the surface of a carbon production target
produced highly polarized positive muons with a momentum of 29.792 MeV/c.
The M13 channel momentum selection was set at 29.6 MeV/c with a resolution of
0.7\% to select these muons and guide them to the spectrometer.
A thin metal foil acted as a stopping target for the muons; this was
placed at the center of a symmetric array of wire chambers
within the bore of a solenoid that produced a highly uniform
$2.0\,\textrm{T}$ magnetic field known to three parts in $10^5$ (see Fig. \ref{f:twist_spectrometer}).
The particle identification relied mostly on three modules of four proportional chambers (PCs) each.
PC modules were installed at each end of the spectrometer. The third module used the stopping target
as a cathode foil in the center of the four PCs and was installed in the center of the detector stack.
Muons that stop in the target were selected using the PCs of this target module.
The decay positron helices were tracked by 44 drift chambers, and
their trajectories were later reconstructed to determine the
positron's initial momentum and angle. The wire positions were known to five parts in $10^5$,
providing a high reconstruction resolution of 60 keV at a positron
energy of 52 MeV. The wire chambers
were low mass to reduce multiple scattering and to allow the muons to reach
the target since it takes only about 1 mm of water equivalent to stop muons at 29 MeV/c.
Further detail on the apparatus can be found elsewhere \cite{Henderson:2005}.

\begin{figure}
\centering
  \includegraphics[height=.3\textheight]{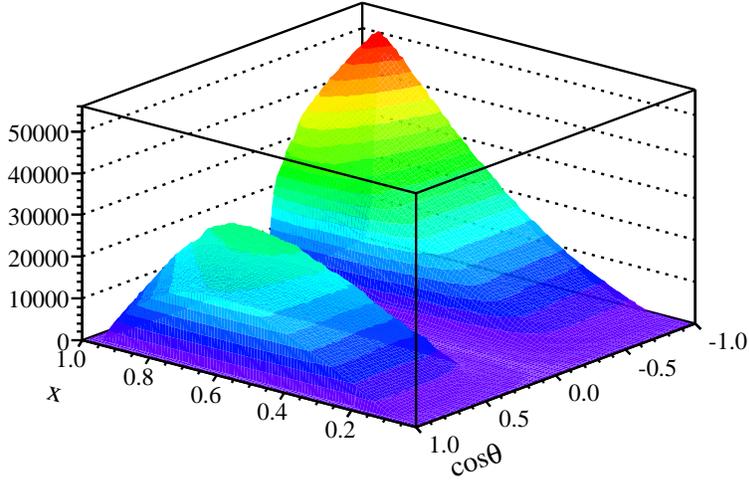}
  \caption{\label{f:twist_spectrum}Experimental momentum-angle spectrum. The detector response is included in this spectrum.}
\end{figure}

The muon decay parameters were measured by comparing the
positron momentum-angle spectra (see Fig. \ref{f:twist_spectrum}) from the data
to a \texttt{GEANT3.21} simulation 
that was subjected to the same analysis.
In this way the detector response and reconstruction biases
were accounted for within the simulation.
Hidden values of $\rho$, $\delta$ and $\xi$ were
used in the simulation, and these were not revealed
until the corrections and systematic uncertainties had
been evaluated on the \emph{difference} in decay
parameters between the data and simulation spectra; this technique
provided a blind analysis by exploiting
the spectrum's linearity in $\rho$, $P_{\mu}^{\pi} \xi$
and $P_{\mu}^{\pi} \xi \delta$ (see Eqs. \eqref{e:Fisotropic},\eqref{e:Fanisotropic}).

Special data validated the positron physics in the
simulation using muons stopped close to the entrance
of the detector. In this configuration the decay positrons traversed
the whole detector. The corresponding tracks were independently reconstructed in
each half of the detector, before and after crossing the stopping target.
The reconstruction efficiency was measured from this special data by counting
the number of tracks reconstructed in one half of the detector but not in the other.

\section{Improvements for the final measurement}
An initial and an intermediate measurement of the decay
parameters were already performed by the TWIST collaboration
on data taken in 2002 \cite{Jim05,Andrei05} and 2004 \cite{Blairpaper,Robpaper}.
Final data were acquired in 2006 and 2007, with 
a higher quality muon beam and a threefold increase
in statistics.

A new technique was developed to measure the space-time relationships (STRs)
used to convert the drift times into drift distances in the drift chambers \cite{Alex:2010}.
These improved STRs also corrected reconstruction biases.
Each drift chamber was calibrated independently;
this accounted for small differences in construction and
response.
The beam line was upgraded to correct an undesirable
muon beam vertical deflection of about 1.0 cm.
The beam was steered onto the symmetry axis of the solenoid,
which reduced the uncertainty in simulating the depolarization of the muon.
The long term stability
of the beam was monitored using its average
position measured by the wire chambers.
Muons were stopped in both an Al and Ag
target foil (previously only an Al foil was used) in two separate datasets.
This allowed for the study of systematics depending on the properties of the target material.

\section{Systematics uncertainties}
The systematic uncertainties from the blind analysis of the final measurement are summarized for each decay parameter in the Table \ref{t:SystUncert}.
\begin{table}
\begin{center}
\caption{\label{t:SystUncert}Systematic uncertainties for each decay parameter in units of $10^{-4}$ from the blind analysis. The depolarization and background muons systematic uncertainties affect only the muon polarization $P_\mu$.}
\begin{tabular}{llll}
\br
									&$\rho$	&$\delta$&$P_\mu^\pi\xi$	\\
\mr
Depolarization in fringe field		&	-	&	-	&$_{-4.0}^{+15.8}$	\\
Depolarization in production target	&	-	&	-	&	3.2			\\
Depolarization in stopping material	&	-	&	-	&	0.3			\\
Background muons					&	-	&	-	&	1.0			\\
Positron interactions				&	1.8	&	1.6	&	0.6			\\
Chamber response					&	1.0	&	1.8	&	2.3			\\
Momentum calibration				&	1.2	&	1.2	&	1.5			\\
Resolution							&	0.6	&	0.7	&	1.5			\\
Alignment							&	0.2	&	0.3	&	0.2			\\
Beam stability						&	0.1	&	0.0	&	0.3			\\
Radiative corrections				&	0.8	&	0.6	&	0.5			\\
Uncertainty in $\eta$				&	1.0	&	0.1	&	1.0			\\
\mr
Total								&	2.8	&	2.9	&$_{-6.2}^{+16.5}$	\\
\br
\end{tabular}
\end{center}
\end{table}

Most systematic uncertainties are evaluated by altering the component source of uncertainty in the simulation or in the analysis. The momentum-angle spectrum created by this modified simulation or analysis is then fitted against the unaltered spectrum.
The alteration is typically many times greater than the measured uncertainty on the modified component in order to increase the sensitivity.
For this reason the difference in decay parameters is scaled down to match the uncertainty on the modified component. This scaled difference is used as the systematic uncertainty on the decay parameters.
For example the bremsstrahlung production rate uncertainty (the dominant uncertainty in the group ``positron interactions'') was evaluated by generating a simulation with the bremsstrahlung production rate multiplied by a factor of 3. The bremsstrahlung rate was measured in the standard simulation and the experimental data using the topology of the events to identify events with a bremsstrahlung being emitted. The simulation and the data bremsstrahlung rates differ by 2.4\%. Therefore the difference in decay parameters between the altered and the unaltered simulations was multiplied by (1.024-1)/(3-1) to provide the corresponding systematic uncertainty on the decay parameters.

The dominant systematic uncertainty for the $P_\mu\xi$ parameter comes from the depolarization undergone by the muons as they enter the 2.0 T tracking magnetic field. A mismatch in the depolarization in the fringe field region between the simulation and the data leads to a mismatch in the muon polarization $P_\mu$ at the time of decay. This creates a systematic bias in the determination of $P_\mu^\pi$ and in the measurement of $P_\mu^\pi\xi$. The accuracy of the simulation of the depolarization was evaluated by modifying the position or size of the experimental muon beam and verifying that the simulation could reproduce the resulting change in polarization. This evaluation indicated that the simulation underestimates the depolarization which leads to an asymmetric uncertainty.

\section{Results}
The TWIST collaboration agreed on the list and values of the systematic uncertainties and corrections before
revealing the hidden parameters of the blind analysis.

The results of the blind analysis are
\begin{eqnarray}
	\rho			&=&	0.74991	\pm 0.00009	\ \mbox{(stat)} \pm 0.00028	\ \mbox{(sys)},\\
	\delta			&=&	0.75072	\pm 0.00016	\ \mbox{(stat)} \pm 0.00029	\ \mbox{(sys)},\\
	P_\mu^\pi\xi	&=&	1.00083	\pm 0.00035	\ \mbox{(stat)} ^{+0.00165}_{-0.00063}	\ \mbox{(sys)}.
\end{eqnarray}

\begin{figure}
\centering
  \includegraphics[width=.65\textwidth]{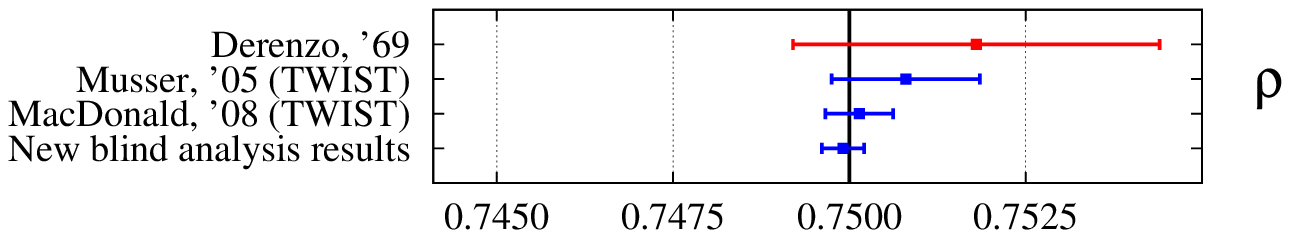}
  \includegraphics[width=.65\textwidth]{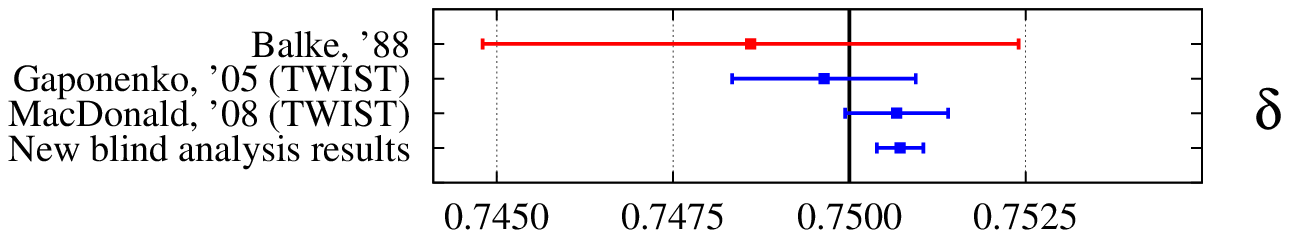}
  \includegraphics[width=.65\textwidth]{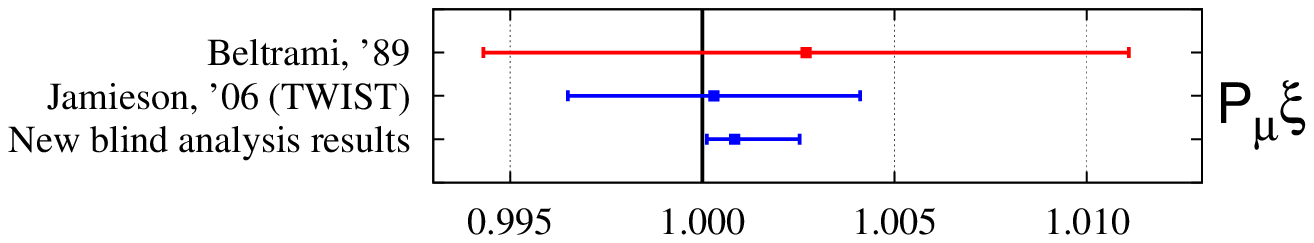}
  \caption{\label{f:twist_consistent}Blind analysis results plotted along with the previous TWIST and pre-TWIST measurements.}
\end{figure}

The parameters $\rho$, $\delta$ and $P_\mu^\pi\xi$ are respectively 0.3, 2.2 and 1.2 standard deviations away from the
predictions of the standard model. All the parameters are consistent with the previous measurements from TWIST
and from experiments prior to TWIST \cite{Derenzo69,Balke:1988,Beltrami:1987} (see Fig. \ref{f:twist_consistent}).

The spectrum asymmetry $A_{EP}$ at the positron kinematic end point
is given by
\begin{equation}
	A_{EP} = \frac{P_\mu^\pi\xi\delta}{\rho},\quad\mbox{therefore }\ \frac{P_\mu^\pi\xi\delta}{\rho} \leq 1.
\end{equation}

However the results from the blind analysis give:
\begin{equation}
	\frac{P_\mu^\pi\xi\delta}{\rho} - 1 = (192 _{-66}^{+167})\times 10^{-5}
\end{equation}
which corresponds to 2.9 standard deviations above the physical limit of one for the four-fermion interaction model.
At the present time we assume that there is a systematic uncertainty or correction that we haven't identified. For
this reason the blind analysis results are not considered final and could be subject to change.

\section{Theoretical implications}
The blind analysis results can be used to put stringent constraints on new physics.
It is important to emphasize that these constraints like the blind analysis results 
given above are not to be considered final.

In Left-Right Symmetric (LRS) models the right-handed current is suppressed but not zero.
An additional
heavy right-handed W-boson ($W_{R}$) is introduced to restore
parity conservation at high energies  \cite{Herczeg:1986}.
In these models, the left- and right-handed gauge boson fields are given by:
\begin{eqnarray}
	W_L &=& W_1 \cos\zeta + W_2 \sin\zeta\\
	W_R &=& e^{i\omega}(-W_1 \sin\zeta + W_2 \cos\zeta)
\end{eqnarray}
where $\omega$ is a CP violating phase.
The TWIST result for $\rho$ allow for model-independent constraints
on the mixing angle $(\zeta)$ between the $W_L$ and $W_R$ and on the mass $m_2$
of the $W_2$ mass eigenstate. No assumptions on the left and right couplings, CKM matrices,
or on the CP violation are made.
The pre-TWIST limits from muon decay were $(g_R/g_L)\vert \zeta \vert < 0.06$ and 
$ (g_R/g_L) m_2 > 400 \mathrm{GeV/c^2}$.
Our preliminary results improve these limits to $(g_R/g_L) \vert \zeta \vert < 0.02$ and $(g_R/g_L) m_2 > 680 \mathrm{GeV/c^2}$.

\section{Conclusions}

The blind analysis of the final data from the TWIST experiment reached the
precision goal of a few $10^{-4}$ on the measurement of the decay parameters
$\rho$, $\delta$ and $P_\mu\xi$. The product $P\mu\xi\delta/\rho$ is 2.9 standard deviations
above the physical limit of one defined by the four-fermion interaction model used.
The present results are therefore not final and the possibility of a missing systematic
uncertainty or correction is being investigated.
\section{Acknowledgments}
This work was supported in part by the Natural Sciences
and Engineering Research Council of Canada, the National
Research Council of Canada, the Russian Ministry of Science,
and the U.S. Department of Energy. Computing resources
for the analysis were provided by WestGrid.

\section*{References}
\bibliographystyle{iopart-num}
\bibliography{LLWIproceedings}

\end{document}